\documentclass[12pt,english]{article}
\usepackage[latin9]{inputenc}
\usepackage{amsmath}
\usepackage{graphicx}
\usepackage{amssymb}
\usepackage{esint}

\makeatletter
\newcommand{\lyxaddress}[1]{
\par {\raggedright #1
\vspace{1.4em}
\noindent\par}
}

\makeatother

\usepackage{babel}

\makeatother

\usepackage{babel}

\begin{document}

\title{On the Properties of the Longitudinal RVB State in the Anisotropic
Triangular Lattice. Mean-Field RVB Analytical Results.}

\author{A.L. Tchougréeff$^{a,b}$ and R. Dronskowski$^{a}$}

\maketitle

\lyxaddress{$^{a}$Institut für anorganische Chemie RWTH-Aachen, Landoltweg 1,
D-52056, Aachen, Germany;\\
 $^{b}$Poncelet Lab., Independent University of Moscow, Moscow
Center for Continuous Mathematical Education, Moscow, Russia. }
\begin{abstract}
The spin-1/2 Heisenberg model on an anisotropic triangular lattice
is considered in the mean-field RVB approximation. The analytical
estimates for the critical temeperatures of the longitudinal s-RVB
state (the upper one) and the 2D s-RVB state (the lower one) are obtained
which fairly agree with the results of the previous numerical studies
on this system. Analytical formulae for the magnetic susceptibility
and the magnetic contribution to the specific heat capacity in the
longitudinal s-RVB state are obtained which fairly reproduce the results
of the numerical experiment concerning these physical quantities. 
\end{abstract}

\section{Introduction}

\begin{figure}
\includegraphics[scale=0.4]{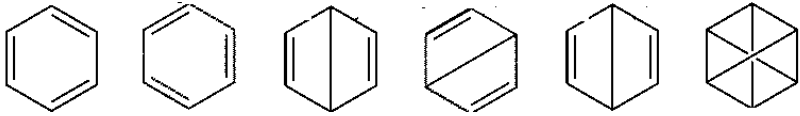} \caption{$\pi$-system of the benzene molecule. From left to right: two Kekulé
resonance structures, three Dewar resonance structures, and the Claus
resonance structure.}

\label{fig:Benzenes} 
\end{figure}

The RVB state originally proposed by Pauling \cite{Pauling} for describing
the structure of benzene molecule as a superposition (resonance) of
its Kekulé \cite{Kekule}, Dewar \cite{Dewar}, and Claus \cite{Claus}
structures as shown in Fig. \ref{fig:Benzenes} %
had been introduced into solid state physics by Anderson \cite{Anderson}
in order to possibly explain the properties of cuprate-based high-temperature
superconductors. Since then this hypothetical ground state is being
sought in many materials. Those exhibiting frustration of the exchange
interactions of the magnetic momenta residing in them are widely suspected
in forming a ground state of the RVB type. Among them the organic
conductors of the family $\kappa$-(BEDT-TTF)$_{2}$Cu$_{2}$XY \cite{fulvalene}
and the halocuprates of the formula Cs$_{2}$CuCl$_{4}$ and Cs$_{2}$CuBr$_{4}$
\cite{halocuplates} are considered as highly propbale candidates.
Recently CuNCN phase had been obtained and a series of measurements
had been performed of its spatial structure and magnetic susceptibility,
electric resistivity, heat capacity (all \emph{vs.} $T$) \cite{Dronskowski}.
Although on the basis of analogy with other materials of the MNCN
series (M = Mn, Fe, Co, Ni) \cite{MnNCN,NiCoNCN,FeNCN} one could
expect more or less standard antiferromagnetic behavior, it turned
out that in the low temperature phase the material at hand does not
manifest any magnetic neutron scattering. The absence of the long
range magnetic order (LRMO) (evanscence of the spin-spin correlation
function) can be explained by the RVB character of the ground state
of the system of the Cu$^{2+}$ local spins 1/2 in this material.
Close inspection of the materials structure (Fig. 2 from Ref. \cite{Dronskowski})%
\begin{figure}
\center{\includegraphics[scale=0.6]{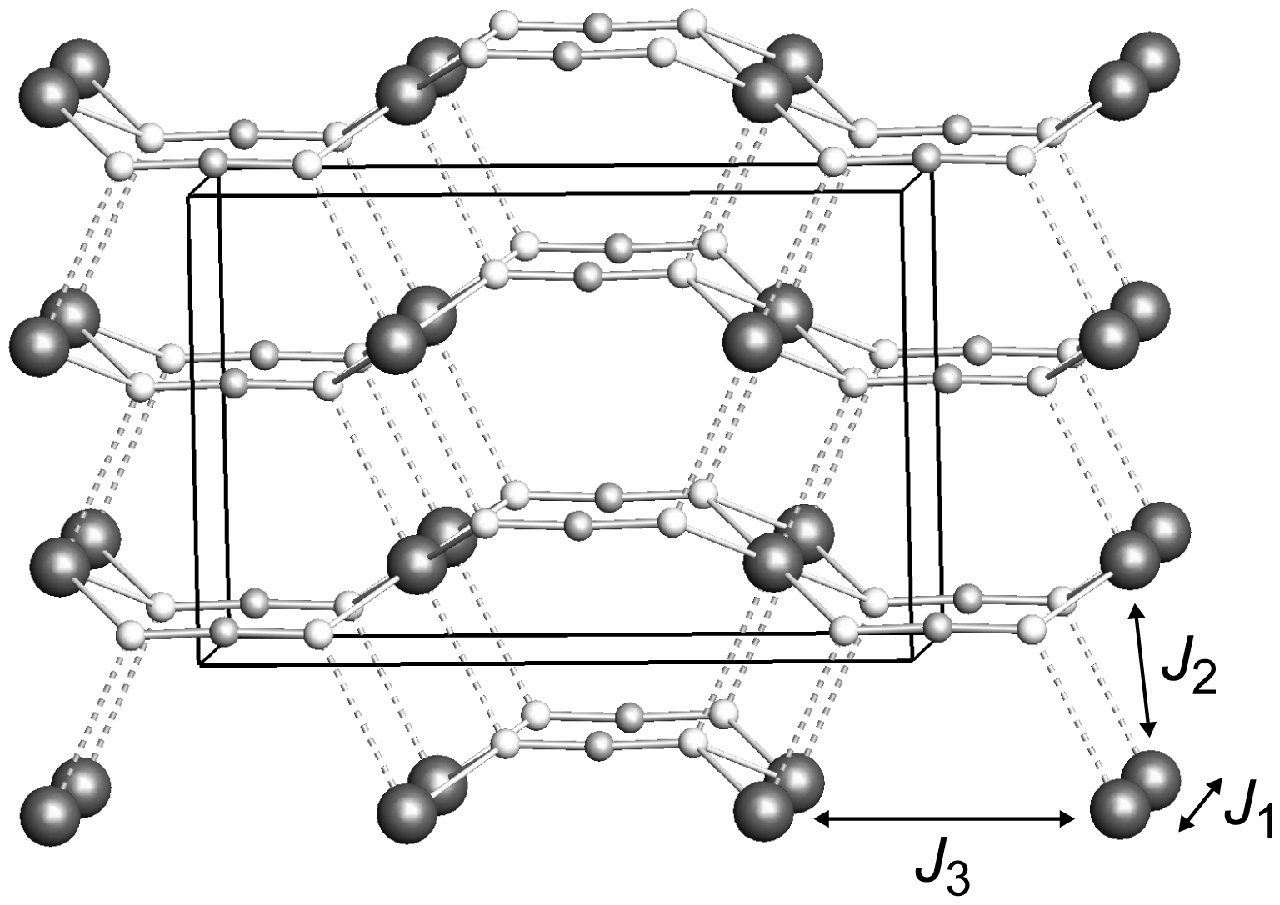}\\
 a\\
 \includegraphics[scale=0.5]{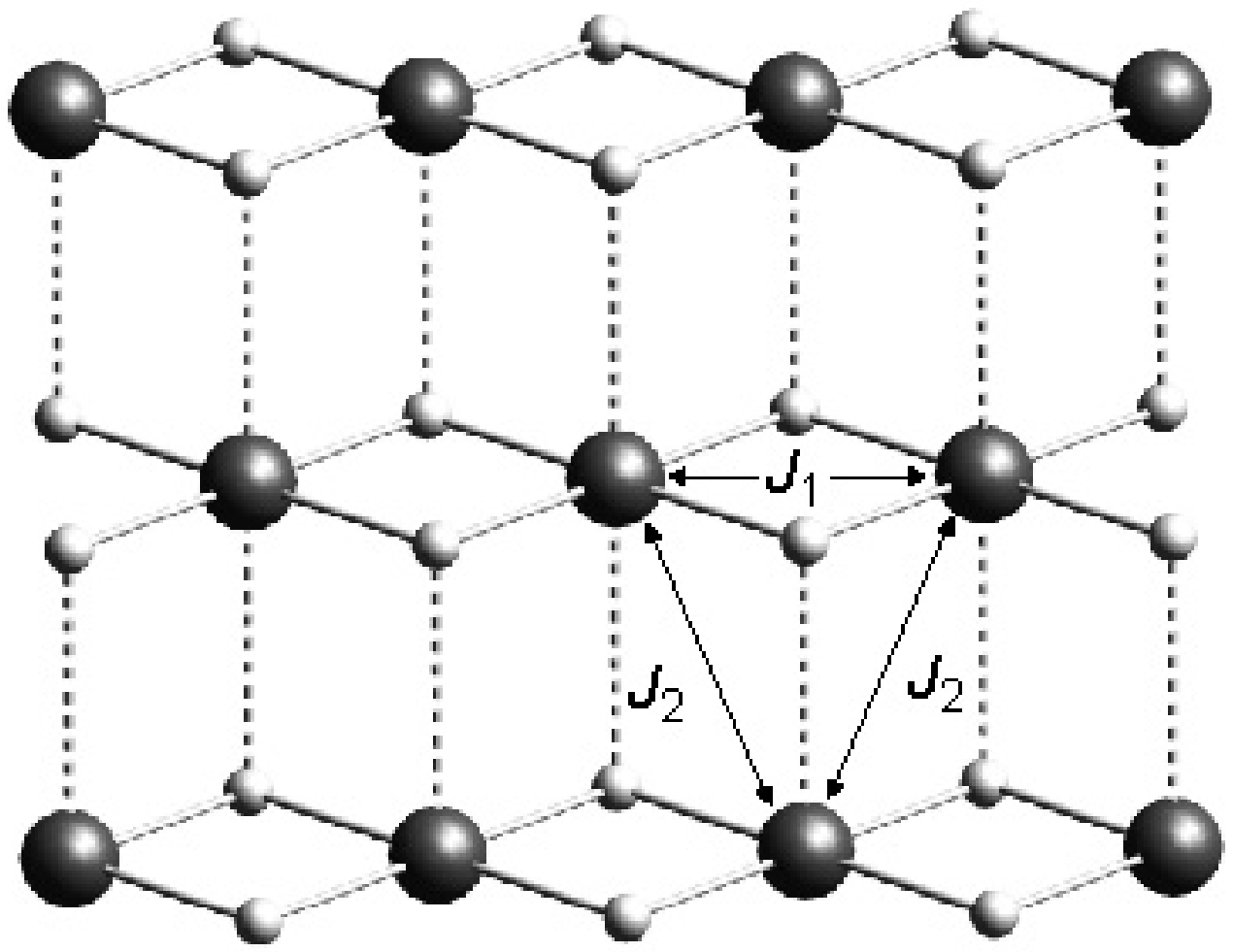}\\
 b} \caption{$ab$ planes of the CuNCN crystal. A stronger $J_{1}$ extends in
the $a$ direction; somewhat weaker $J_{2}$ extends along the $b\pm a$
directions. The weakest $J_{3}$ extends in the $c$ direction and
is not considered in the present paper.}

\label{fig:CuNCN-ab-layer} 
\end{figure}
reveals that each Cu$^{2+}$ ion can be effectively antiferromagnetically
coupled to two its neighbors forming a chain while somewhat weaker
antiferromagnetic coupling with four more neighbours from two adjacent
parallel chains results in a Heisenberg model on an anisotropic triangular
lattice with the Hamiltonian: \begin{equation}
\sum_{\mathbf{r}}\sum_{\mathbf{\tau}}J_{\mathbf{\boldsymbol{\tau}}}\mathbf{S}_{\mathbf{r}}\mathbf{S}_{\mathbf{r}+\tau}\label{eq:Hamiltonian}\end{equation}
where the coupling vectors $\mathbf{\tau}$ take three values $\tau_{i};\, i=1\div3;\,\tau_{1}=(1,0);\,\tau_{2}=(\frac{1}{2},\frac{\sqrt{3}}{2});\,\tau_{3}=(\frac{1}{2},-\frac{\sqrt{3}}{2})$
with the interaction of the strength $J$ along the lattice vector
$\mathbf{\mathbf{\tau}_{\mathrm{1}}}$ (two neighbors) and with a
somewhat smaller strength $J^{\prime}$ along the lattice vectors
$\mathbf{\mathbf{\tau}_{\mathrm{2}}}$ and $\mathbf{\mathbf{\tau}_{\mathrm{3}}}$
(two neighbors along each). This is precisely the setting for which
Hayashi and Ogata proposed that two (different) spin-singlet RVB (s-RVB)
states are formed at different temperatures depending on the amount
of anisotropy $\frac{J^{\prime}}{J}$ resulting in a temperature dependence
of the magnetic susceptibility with characteristic discontinuities
corresponding to installment of these two respective states \cite{Hayashi and Ogata }.

In the present paper we reasses their numerical results analytically
partially with use of the Ginzburg-Landau phase transition theory
in order to further apply the obtained analytical formulae to the
experimental data obtained for CuNCN \cite{Dronskowski}.

\section{RVB mean-field analysis of anisotropic triangular lattice system}

\subsection{Equations of motion and self consistency equations}

Hayashi and Ogata \cite{Hayashi and Ogata } base their analysis of
the Hamiltonian eq. (\ref{eq:Hamiltonian}) on returning to the electron
representation from the spin representation by the standard formulae:\begin{equation}
\mathbf{S}_{i}=\frac{1}{2}c_{i\alpha}^{+}\mathbf{\boldsymbol{\sigma}}_{\alpha\beta}c_{i\beta},\label{eq:SpinThroughFermi}\end{equation}
 where $c_{i\sigma}^{+}(c_{i\sigma})$ are the electron creation (annihilation)
operators subject to the Fermi anticommutation relations; $\mathbf{\boldsymbol{\sigma}}_{\alpha\beta}$
are the elements of the Pauli matrices and the summation over repeating
indices is assumed. For the latter one can derive equations of motion
based on the Heisenberg representation in which each operator obeys
the following equation of motion:\begin{equation}
i\hbar\dot{A}=\left[A,H\right]\label{eq:HeisenbergEOM}\end{equation}
 where $\left[,\right]$ stands for the commutator of the operators
and \char`\"{}$\dot{}$\char`\"{} for the time derivative. Applying
this to the creation and annihilation operators $c_{\mathbf{r}\sigma}^{+}(c_{\mathbf{r}\sigma})$
and performing commutation, mean field decoupling and Fourier transformation
as done in Appendix \ref{sec:Derivation-of-equations} results in
mean field equations of motion for these operators : \begin{eqnarray}
i\hbar\dot{c}_{\mathbf{k}\sigma}=-\frac{3}{2}\sum_{\mathbf{\tau}}J_{\mathbf{\boldsymbol{\tau}}}\xi_{\mathbf{\boldsymbol{\tau}}}\cos(\mathbf{k\tau})c_{\mathbf{k}\sigma} & -\frac{3}{2}\sum_{\mathbf{\tau}}J_{\mathbf{\boldsymbol{\tau}}}\Delta_{\boldsymbol{\tau}} & \cos(\mathbf{k\tau})c_{-\mathbf{k}-\sigma}^{+}\nonumber \\
i\hbar\dot{c}_{\mathbf{k}\sigma}^{+}=\frac{3}{2}\sum_{\mathbf{\tau}}J_{\mathbf{\boldsymbol{\tau}}}\xi_{\mathbf{\boldsymbol{\tau}}}\cos(\mathbf{k\tau})c_{\mathbf{k}\beta}^{+} & +\frac{3}{2}\sum_{\mathbf{\tau}}J_{\mathbf{\boldsymbol{\tau}}}\Delta_{\boldsymbol{\tau}}^{*} & \cos(\mathbf{k\tau})c_{-\mathbf{k}-\sigma}\label{eq:MeanFieldEOM}\end{eqnarray}
These latter reduces to the set of $2\times2$ eigenvalue problems
for each wave vector $\boldsymbol{\mathbf{k}}$:

\[
\left(\begin{array}{cc}
\xi_{\mathbf{k}} & \Delta_{\mathbf{k}}\\
\Delta_{\mathbf{k}}^{*} & -\xi_{\mathbf{k}}\end{array}\right)\left(\begin{array}{c}
u_{\mathbf{k}}\\
v_{\mathbf{k}}\end{array}\right)=E_{\mathbf{k}}\left(\begin{array}{c}
u_{\mathbf{k}}\\
v_{\mathbf{k}}\end{array}\right)\]
with\begin{eqnarray}
\xi_{\mathbf{k}} & = & -3\sum_{\mathbf{\tau}}J_{\mathbf{\boldsymbol{\tau}}}\xi_{\mathbf{\boldsymbol{\tau}}}\cos(\mathbf{k\tau})\nonumber \\
\Delta_{\mathbf{k}} & = & 3\sum_{\mathbf{\tau}}J_{\mathbf{\boldsymbol{\tau}}}\Delta_{\mathbf{\boldsymbol{\tau}}}\cos(\mathbf{k\tau})\label{eq:DispersionFunctions}\end{eqnarray}
(summation over $\mathbf{\boldsymbol{\tau}}$ extends to $\pm\tau_{i};i=1\div3$)
which results in the eigenvalues (excitation spectrum) of the form:

\begin{eqnarray}
E_{\mathbf{k}} & = & \sqrt{\xi_{\mathbf{k}}^{2}+\left|\Delta_{\mathbf{k}}\right|^{2}}\label{eq:eigenvalues}\end{eqnarray}
whose eigenvectors are combinations of the destruction and creation
operators with the above Bogoliubov transformation coefficients $u_{\mathbf{k}},v_{\mathbf{k}}$.
These equations result in the selfconsistency equations of the form:\begin{eqnarray}
\xi_{\mathbf{\boldsymbol{\tau}}} & = & -\frac{1}{2N}\sum_{\mathbf{k}}\exp(i\mathbf{k\tau})\frac{\xi_{\mathbf{k}}}{E_{\mathbf{k}}}\tanh\left(\frac{E_{\mathbf{k}}}{2\theta}\right)\nonumber \\
\Delta_{\mathbf{\boldsymbol{\tau}}} & = & \frac{1}{2N}\sum_{\mathbf{k}}\exp(-i\mathbf{k\tau})\frac{\Delta_{\mathbf{k}}}{E_{\mathbf{k}}}\tanh\left(\frac{E_{\mathbf{k}}}{2\theta}\right)\label{eq:SelfConsistencyEquations}\end{eqnarray}
for six order parameters $\xi_{\mathbf{\boldsymbol{\tau}}},\Delta_{\mathbf{\boldsymbol{\tau}}}$.

\subsection{Free energy}

Following Ref. \cite{Ogata&Fukuyama} one can write immeditaly the
free energy in terms of the above order parameters: \begin{equation}
-\frac{\theta}{2N}\sum_{\mathbf{k}}\ln\left(2\cosh\left(\frac{E_{\mathbf{k}}}{2\theta}\right)\right)+\frac{3}{2}\sum_{\mathbf{\tau}}J_{\mathbf{\boldsymbol{\tau}}}\xi_{\mathbf{\boldsymbol{\tau}}}^{2}+\frac{3}{2}\sum_{\mathbf{\tau}}J_{\mathbf{\boldsymbol{\tau}}}\left|\Delta_{\mathbf{\mathbf{\boldsymbol{\tau}}}}\right|^{2}\label{eq:FreeEnergy}\end{equation}
(summation over $\mathbf{\boldsymbol{\tau}}$ extends to $\pm\tau_{i};i=1\div3$).
Minima of this expression correspond to various possible states of
the system. This is more informative than studying the self consistency
equations eq. (\ref{eq:SelfConsistencyEquations}) which may have
more solutions. \emph{E.g.} the trivial solution \[
\xi_{\mathbf{\boldsymbol{\tau}}}=\Delta_{\mathbf{\boldsymbol{\tau}}}=0\]
always exist although may be not of the lowest energy (see below).

\section{Simplified model of s-RVB on the anisotropic triangular lattice}

\noindent The numerical analysis of Ref. \cite{Hayashi and Ogata }
shows that in agreement with general theorems Ref. \cite{SU2-symmetry}
the order parameters satisfy additional phase relations (see Appendix
\ref{sub:Phase-relations-for} for details) which allows to reduce
the number of order parameters to only two:\[
\sqrt{2}\xi=\xi_{\boldsymbol{\tau}_{1}};\sqrt{2}\eta=\left|\Delta_{\boldsymbol{\tau}_{2}}\right|=\left|\Delta_{\boldsymbol{\tau}_{3}}\right|\]
the first responsible for establishing the s-RVB state within the
chains (longitudinal s-RVB) and the second for the same in the transversal
direction (2D s-RVB). For the triangular anisotropic lattice as represented
in the above approximation one can easily write the explicit expression
for the free energy as relying on the general expression eq. (\ref{eq:FreeEnergy}).
It reads as follows:

\begin{equation}
F=6J\xi^{2}+12J^{\prime}\eta^{2}-\frac{\theta\sqrt{3}}{4\pi^{2}}\intop_{BZ}\ln\left(2\cosh\left(\frac{E_{\mathbf{k}}}{2\theta}\right)\right)d^{2}\mathbf{k},\label{eq:FreeEnergySimplified}\end{equation}
where $\theta=k_{B}T$ and \begin{eqnarray}
E_{\mathbf{k}}^{2} & = & 18\left[J^{2}\xi^{2}\cos^{2}(\mathbf{k}_{x})+J^{\prime2}\eta^{2}\left(\cos^{2}(\frac{\mathbf{k}_{x}}{2}+\frac{\mathbf{k}_{y}\sqrt{3}}{2})+\cos^{2}(\frac{\mathbf{k}_{x}}{2}-\frac{\mathbf{k}_{y}\sqrt{3}}{2})\right)\right].\label{eq:Spectrum}\end{eqnarray}

It is easy to check that for $\theta\rightarrow0$ the integral produces
an expression which is uniform with respect to $\xi$ and $\eta$
so that the minimum of the free energy corresponds to both order parameters
nonvanishing (see also below):\[
\xi(\theta\rightarrow0)\neq0;\eta(\theta\rightarrow0)\neq0.\]
 At high temperature we can use a high-temperature expansion: \[
\ln\left(2\cosh\left(\frac{E_{\mathbf{k}}}{2\theta}\right)\right)\approx\ln2+\frac{1}{2}\left(\frac{E_{\mathbf{k}}}{2\theta}\right)^{2}-\frac{1}{12}\left(\frac{E_{\mathbf{k}}}{2\theta}\right)^{4}\]
which allows to be integrated explicitly. This results in the Ginzburg-Landau
approximate free energy $F_{\mathrm{GL}}(\xi,\eta,\theta)$ which
is of the fourth power in $\xi$ and $\eta$. The critical temperatures
in this approximation correspond to those where the free energy $F_{\mathrm{GL}}$
of a state with one or two nonvanishing order parameters becomes lower
than that of some other state (\emph{e.g.} of the paramagnetic state
with vanishing order parameters). It happens at:

\begin{equation}
\theta_{c}=\frac{3}{8}J,\label{eq:UpperCriticalTemperature}\end{equation}
where the longitudinal s-RVB state with:

\begin{equation}
\xi=\frac{2\sqrt{2}\theta\sqrt{1-\theta/\theta_{c}}}{3J};\eta=0\label{eq:OrderParametersLongitudinalRVB}\end{equation}
installs (in perfect correspondence to the analytical result announced
in Ref. \cite{Hayashi and Ogata }) and at \begin{equation}
\theta_{c}^{\prime}=\frac{3JJ^{\prime}}{8(3J-2J^{\prime})}\label{eq:LowerCriticalTemperature}\end{equation}
 with \[
\frac{\theta_{c}}{\theta_{c}^{\prime}}=\frac{3J}{J^{\prime}}-2>1.\]
Below the lower critical temperature eq. (\ref{eq:LowerCriticalTemperature})
the 2D-RVB state installs with the order parameters:\begin{eqnarray}
\xi & = & \frac{2\sqrt{2}}{3}\frac{\theta}{J}\sqrt{\left(1-\frac{\theta}{\theta_{c}}\right)-\frac{4}{7}\left(1-\frac{\theta}{\theta_{c}^{\prime}}\right)}\nonumber \\
\eta & = & \sqrt{\frac{8}{21}}\frac{\theta}{J^{\prime}}\sqrt{1-\frac{\theta}{\theta_{c}^{\prime}}}\label{eq:OrderParameters2D-RVB}\end{eqnarray}
which is in fair agreement with the numerical result of Ref. \cite{Hayashi and Ogata }
in that sence that it shows some depletion of the longitudinal order
parameter ($\xi$) below the critical temperature of the transition
to the 2D-RVB state. The order parameters' estimates are obtained
as solutions for the minimum conditions for the Landau-Ginzburg free
energy $\partial F_{\mathrm{GL}}(\xi,\eta,\theta)/\partial\xi=\partial F_{\mathrm{GL}}(\xi,\eta,\theta)/\partial\eta=0$
as functions of $\theta$. The lower critical temperature eq. (\ref{eq:LowerCriticalTemperature})
is, as one can expect, an overestimate of the exact critical temperature
usual for the mean field theory as combined with the high-temperature
expansion. It is, nevertheless, in a reasonable agreement with the
numerical results of Ref. \cite{Hayashi and Ogata } where $\theta_{c}^{\prime}/\theta_{c}$
was estimated to be 0.213 for $\frac{J^{\prime}}{J}=0.6$ whereas
our estimate is $1/3$ for this amount of anisotropy.

For the longitudinal s-RVB phase eq. (\ref{eq:OrderParametersLongitudinalRVB})
further analysis is possible. The integral term in the limit $\theta\rightarrow0$
reduces to \[
-\frac{1}{\pi}\int_{-\frac{\pi}{2}}^{\frac{\pi}{2}}E_{\mathbf{k}}d\mathbf{k}_{x}=-\frac{6\sqrt{2}}{\pi}J\xi\]
with use of the integration method as described in Appendix \ref{sec:Integration-in-the}.
As combined with the term squared in $\xi$ in the free energy eq.
(\ref{eq:FreeEnergySimplified}) this yields the amplitude of the
order parameter $\xi$ reached at the zero temperature:\begin{equation}
\xi_{0}=\frac{1}{\sqrt{2}\pi},\label{eq:LimitOrderParameter}\end{equation}
which is in perfect agreement with the numerical result of Ref. \cite{Hayashi and Ogata }.

\section{Physical properties of the anisotropic triangular Heisenberg model}

Qualitatively the behavior of the anisotropic triangular Heisenberg
model looks as follows: order parameters related to the intrachain
and interchain interactions are both vanishing above the higher critical
temperature $\theta_{c}$. The excitation spectrum is gapless and
there is no dispersion in either direction so that paramagnetic behavior
is to be expected (see below). Below the first critical temperature
(since $\frac{J^{\prime}}{J}<1$ ) $\eta=0$ (longitudinal s-RVB state)
the excitations have the energy spectrum:\begin{eqnarray}
E_{\mathbf{k}} & =3\sqrt{2} & \left|J\xi\cos(\mathbf{k}_{x})\right|\label{eq:LongitudinalSpectrum}\end{eqnarray}
with no dispersion in the transversal direction although with the
temperature dependent bandwidth. The excitations propagating with
the wave vectors with the longitudinal components equal to $\pm\frac{\pi}{2}$
are gapless. Below the lower critical temperature one can expect that
provided the ratio $\frac{J^{\prime}}{J}$ is not too large the spectrum
acquires a finite gap for whatever wave vector. Due to qualitative
changes in the spectrum of the system occurring at the two critical
temperatures one can expect that the physical quantities which are
predominantly contributed by the excitations with low energies will
be significanly different in different phases. In what follows we
shall derive analytical formulae for the magnetic contribution to
the specific heat capacity and the magnetic susceptibility in the
longitudinal s-RVB state (between two critical temperatures eqs. (\ref{eq:UpperCriticalTemperature}),
(\ref{eq:LowerCriticalTemperature})) where only the order parameter
$\xi$ is nonvanishing. In the temperature range of interest the properties
are predetermined by the vicinity of the gapless excitations along
the lines $\mathbf{k}_{x}=\pm\frac{\pi}{2}$ in the Brillouin zone.
However, the detailed form the spectrum must be immaterial for obtaining
some semiquantitative estimates provided the qualitative features
of the spectrum are captured. This can be done with use of a saw-like
model dispersion law: \begin{eqnarray}
E_{\mathbf{k}} & =3\sqrt{2} & \left|J\xi(1-\frac{2\left|\mathbf{k}_{x}\right|}{\pi})\right|,\label{eq:SawSpectrum}\end{eqnarray}
which has the same gap behavior and the same bandwidth as the exact
problem. It corresponds to the constant density of quasiparticle states
in variance with the exact spectrum which is characterized by accumulation
of the density at the top of the quasiparticle band.

\subsection{Specific heat capacity}

The magnetic contribution to the specific heat derives using the standard
definition:

\[
C_{m}=-2k_{B}T\frac{\partial}{\partial T}\sum_{\mathbf{k}}\left[f(E_{\mathbf{k}})\ln f(E_{\mathbf{k}})+(1-f(E_{\mathbf{k}}))\ln(1-f(E_{\mathbf{k}}))\right],\]
 where \[
f(E)=\left[\exp(\frac{E}{\theta})+1\right]^{-1}\]
 is the Fermi distribution function. The specific heat capacity apparenly
vanishes in the gapless state (above $\theta_{c}$). Replacing the
summation by integration over BZ as described in Appendix \ref{sec:Integration-in-the}
it becomes one per individual spin. After taking derivative with respect
to $T$ it becomes:\[
C_{m}=k_{B}\frac{\sqrt{3}}{16\pi^{2}}\int_{BZ}\frac{E_{\mathbf{k}}\left(E_{\mathbf{k}}-TE_{\mathbf{k}}^{\prime}\xi^{\prime}\right)\text{sech}^{2}\left(\frac{E_{\mathbf{k}}}{2\theta}\right)}{\theta^{2}}d^{2}\mathbf{k}.\]
where $E_{\mathbf{k}}^{\prime}$ stands for the derivative of $E_{\mathbf{k}}$
with respect to $\xi$ and $\xi{}^{\prime}$ stands for derivative
of $\xi$ with respect to $T$.

\subsubsection{Saw-like spectrum}

For the saw-like spectrum in the area where the integration is to
be performed the following

\[
E_{\mathbf{k}}\left(E_{\mathbf{k}}-TE_{\mathbf{k}}^{\prime}\xi^{\prime}\right)=18J^{2}(1-\frac{2\mathbf{k}_{x}}{\pi})^{2}\xi\left(\xi-T\xi^{\prime}\right),\]
 holds, so that \[
C_{m}=k_{B}\frac{y^{2}\left(1-T(\ln y)^{\prime}\right)}{\pi\theta^{2}}\int_{0}^{\frac{\pi}{2}}(1-\frac{2x}{\pi})^{2}\text{sech}^{2}\left(\frac{y(1-\frac{2x}{\pi})}{2\theta}\right)dx,\]
 where $y=2\sqrt{3}J\xi$. The integration can be performed explicitly,
which yields:\begin{equation}
\left(\frac{\tanh\left(\frac{y}{2\theta}\right)}{\theta}-\frac{1}{\theta}-\frac{4}{y}\ln\left(1+e^{-\frac{y}{\theta}}\right)+\frac{\pi^{2}\theta}{3y^{2}}+\frac{4\theta\text{Li}_{2}\left(-e^{-\frac{y}{\theta}}\right)}{y^{2}}\right)y\left(1-T(\ln y)^{\prime}\right)\label{eq:HeatCapacityLongitudinalSawLike}\end{equation}
For the divergent $\frac{y}{\theta}$ one can expand hyprbolic tangent,
the logarithm, and polylogarithm, and make sure that the divergent
terms cancel each other and that the dominant contribution to the
specific heat capacity is linear in temperature: \[
C_{m}=\frac{\pi^{2}\theta}{3y}\left(1-T(\ln y)^{\prime}\right)k_{B}=\frac{\pi^{2}\theta}{9\sqrt{2}J\xi}k_{B}\]
(correction to it is at best proportional to $\frac{y}{\theta}e^{-\frac{y}{\theta}}$
which clearly disappears for the divergent $\frac{y}{\theta}$). On
the other end of the interval -- at the critical temperature $\theta_{c}$,
by contract, $\frac{\theta}{y}$ diverges according to the divergence
law of $\xi$. Expanding the first bracket in the above expression
eq. (\ref{eq:HeatCapacityLongitudinalSawLike}) at small values of
$\xi$ we obtain: \begin{equation}
C_{m}=k_{B}\frac{3J^{2}\xi^{2}}{\theta^{2}}\left(1-T(\ln\xi)^{\prime}\right).\label{eq:HeatCapacity}\end{equation}
This exact result can be combined with the classical expression for
the temperature dependence of the order parameter in the vicinity
of the critical point (for the reasons which will be clear below we
use here a generalized form of the classical temperature dependency
of the order parameter $(1-\theta/\theta_{c})^{1/2}$ which shows
the same critical exponent as the classical one):\begin{eqnarray}
\xi & = & \xi_{0}\sqrt{1-\left(\theta/\theta_{c}\right)^{\alpha}}\nonumber \\
T\xi^{\prime} & =-\frac{\alpha}{2}\xi_{0}\left(\frac{\theta}{\theta_{c}}\right)^{\alpha} & \frac{1}{\sqrt{1-\left(\theta/\theta_{c}\right)^{\alpha}}}\label{eq:GeneralizedClassical}\end{eqnarray}
First term in the brackets eq. (\ref{eq:HeatCapacity}) when multiplied
by $\xi^{2}$ is small in a vicinity of $\theta_{c}$. The second
term yields:\[
-\frac{T\xi^{\prime}}{\xi}=\frac{\alpha}{2}\left(\frac{\theta}{\theta_{c}}\right)^{\alpha}\frac{1}{1-\left(\theta/\theta_{c}\right)^{\alpha}}=\frac{\alpha}{2}\left(\frac{\theta}{\theta_{c}}\right)^{\alpha}\left(\frac{\xi_{0}}{\xi}\right)^{2}\]
so that, combining this with eq. (\ref{eq:HeatCapacity}) we obtain
the finite jump of the magnetic contribution to the specific heat
capacity in the critical point: \begin{equation}
\Delta C_{m}(\theta_{c})=k_{B}\frac{3J^{2}\alpha}{\theta_{c}^{2}}\xi_{0}^{2}=\frac{16\alpha}{3\pi^{2}}k_{B}.\label{eq:SawHeatCapacityJump}\end{equation}
This finite jump is qualitatively in agreement with general theory.
Precise agreement with the numerical resutls of \cite{Hayashi and Ogata }
is not achieved due to approximate form both of the spectrum and the
temperature dependence of the order parameter used above. On the other
hand required agreement can be achieved by fitting the value of $\alpha$
(see below).

\subsubsection{Exact spectrum}

If the exact spectrum is inserted in the general expression for the
magnetic contribution to the specific heat capacity we get:

\[
C_{m}=k_{B}\frac{y^{2}\left(1-T(\ln y)^{\prime}\right)}{\pi\theta^{2}}\int_{0}^{\frac{\pi}{2}}\cos{}^{2}x\text{ sech}^{2}\left(\frac{y\cos x}{2\theta}\right)dx,\]
We could not find analytical expression for this integral. However,
the magnitude of the jump of the magnetic contribution to the specific
heat capacity in the critical point can be estimated from an expansion
of the above integral for small $y\,(\xi)$. The result in the vicinity
of the critical point reads:\[
C_{m}=k_{B}\frac{9J^{2}\xi^{2}}{2\theta^{2}}\left(1-T(\ln\xi)^{\prime}\right).\]
Combining this with the above expression for the logarithmic derivative
of the order parameter we obtain:\begin{equation}
\Delta C_{m}(\theta_{c})=\frac{8\alpha}{\pi^{2}}k_{B}.\label{eq:ExactHeatCapacityJump}\end{equation}
Analysis of the low temperature limit for the exact heat capacity
will be performed elsewhere.

\subsection{Magnetic susceptibility}

We use the standard definition of the magnetic susceptibily per spin:
\[
\chi=-2\mu_{B}^{2}\frac{1}{N}\sum_{\mathbf{k}}\frac{\partial f(E_{\mathbf{k}})}{\partial E_{\mathbf{k}}}\]
The summation replaces by integration which yields: \begin{equation}
\chi=-\frac{\mu_{B}^{2}\sqrt{3}}{4\pi^{2}}\intop_{BZ}\frac{\partial f(E_{\mathbf{k}})}{\partial E_{\mathbf{k}}}d^{2}\mathbf{k}\label{eq:SusceptibilityDefinition}\end{equation}
Following the recipe of Appendix \ref{sub:Using-geometric-series}
we get for the susceptibility: \[
\chi=\frac{\mu_{B}^{2}\sqrt{3}}{4\pi^{2}\theta}\sum_{n=1}^{\infty}\left(-1\right)^{n+1}n\intop_{BZ}\exp\left(-\frac{nE_{\mathbf{k}}}{\theta}\right)d^{2}\mathbf{k},\]
which is analyzed below.

\subsubsection{High temperature ($\theta>\theta_{c}$) gapless state}

In this state the argument of the function under integral equals to
zero for all values of $\mathbf{k}$. Thus inserting $E_{\mathbf{k}}\equiv0$
to the derivative of the Fermi distirbution function\begin{equation}
\frac{\partial f(E)}{\partial E}=-\frac{\exp(\frac{E}{\theta})}{\theta\left[\exp(\frac{E}{\theta})+1\right]^{2}}\label{eq:FermiDistributionDerivative}\end{equation}
we obtain:\[
\chi=\frac{\mu_{B}^{2}}{2\theta};\theta>\theta_{c}\]
in perfect agreement with the numerical result reported by Hayashi
and Ogata \cite{Hayashi and Ogata }.

\subsubsection{Longitudinal s-RVB state}

In this state the gap is absent along the lines $\mathbf{k}_{x}=\pm\frac{\pi}{2}$
in the reciprocal space. Their vicinity provides major contribution
to the susceptibility. It will be estimanted below for the saw-like
approximation of the spectrum and for the exact one.

\paragraph{Saw like approximation of the spectrum}

Performing the integration as described in Appendix \ref{sec:Integration-in-the}
we get for the susceptibility: \[
\chi=\frac{\mu_{B}^{2}}{J\xi}\frac{\sqrt{2}}{3}\sum_{n=1}^{\infty}\left(-1\right)^{n+1}\left(1-\exp\left(-\frac{3\sqrt{2}J\xi n}{\theta}\right)\right).\]
Summation of the constant term yields $\frac{1}{2}$ according to
Appendix \ref{sub:Using-geometric-series} and the rest is just the
geometric series so that the susceptibility becomes:\[
\chi=\frac{\mu_{B}^{2}}{J\xi}\frac{\sqrt{2}}{3}\left(\frac{1}{2}-\frac{\exp\left(-\frac{3\sqrt{2}J\xi}{\theta}\right)}{1+\exp\left(-\frac{3\sqrt{2}J\xi}{\theta}\right)}\right),\]
 which transforms into\[
\chi=\frac{\mu_{B}^{2}}{J\xi}\frac{1}{3\sqrt{2}}\tanh\left(\frac{3J\xi}{\sqrt{2}\theta}\right).\]

\paragraph{Qualitative behavior of the saw-spectrum model susceptibility}

The temperature behavior of the susceptibility in the saw approximation
depends on that of $\xi(\theta)$. If the latter flows to a finite
value while $\theta\rightarrow0$ the susceptibily remains finite
as well in agreement with finding of Ref. \cite{Hayashi and Ogata }
for the longitudinal state. The limiting value equals $\frac{\pi}{3}\frac{\mu_{B}^{2}}{J}$
which is significantly higher than the numerical limiting value obtained
in Ref. \cite{Hayashi and Ogata }. By contrast, if $\frac{\xi(\theta)}{\theta}\rightarrow\alpha>0$
(the case of the temperature dependence provided by the high-temperature
expansion eq. (\ref{eq:OrderParametersLongitudinalRVB})), a paramagnetic
behavior can be expected, although with the number of momenta effectively
suppressed by hyperbolic tangens which is always less than unity.

\paragraph{Exact susceptibility in the longitudinal s-RVB state}

Using the integration trick as described in Appendix \ref{sec:Integration-in-the}
and inserting the exact spectrum of the longitudinal state and noticing
the argument of the absolute value is positive in the integration
limits we obtain for the integral:\begin{eqnarray*}
\int_{-\frac{\pi}{2}}^{\frac{\pi}{2}}\exp\left(-\frac{nE_{\mathbf{k}}}{\theta}\right)d\mathbf{k}_{x} & = & \int_{-\frac{\pi}{2}}^{\frac{\pi}{2}}\exp\left(-\frac{n3\sqrt{2}J\xi\cos(\mathbf{k}_{x})}{\theta}\right)d\mathbf{k}_{x}\\
 & = & \pi\left(I_{0}\left(\frac{n3\sqrt{2}J\xi}{\theta}\right)-L_{0}\left(\frac{n3\sqrt{2}J\xi}{\theta}\right)\right)\end{eqnarray*}
where $I_{0}$ is the modified Bessel function and $L_{0}$ is the
Struve function of the 0-th index. So that we arrive to the expression
for the susceptibility in the longitudinal s-RVB state

\begin{equation}
\chi(\theta)=\frac{2\mu_{B}^{2}}{\theta}\sum_{n=1}^{\infty}\left(-1\right)^{n+1}n\left(I_{0}\left(\frac{n3\sqrt{2}J\xi}{\theta}\right)-L_{0}\left(\frac{n3\sqrt{2}J\xi}{\theta}\right)\right),\label{eq:SusceptibilyExact}\end{equation}
which is exact.

\paragraph{Qualitative behavior of the exact susceptibility}

The differences of the modified Bessel function and the Struve function
in eq. (\ref{eq:SusceptibilyExact}) divided by $\theta$ have finite
limit while $\theta\rightarrow0$ which equals to \[
\frac{\sqrt{2}}{3\pi J\xi n}.\]
Inserting it into the series eq. (\ref{eq:SusceptibilyExact}) and
performing summation according to Appendix \ref{sec:Convergence-issues}
we obtain that at zero temperature the susceptibility in the longitudinal
s-RVB state has finite limit \[
\chi=\frac{\mu_{B}^{2}\sqrt{2}}{3\pi J\xi_{0}}=\frac{2\mu_{B}^{2}}{3J}\]
This is in perfect agreement with the numerical result of Ref. \cite{Hayashi and Ogata }.
The ratio \[
\frac{\chi^{\mathrm{exact}}(\theta=0)}{\chi^{\mathrm{saw}}(\theta=0)}=\frac{2}{\pi}<1\]
which is expectable due to accumulation of the quasiparticle states
at the top of the exact spectrum \emph{vs} uniform density of states
for the saw-like spectrum. On the other hand in the vicinity of the
critial temperature $\theta_{c}$ where order parameter is small in
eq. (\ref{eq:SusceptibilyExact}) the series terms have finite limit
of unity independently of the $n$ value, which as combined with Appendix
\ref{sub:Using-geometric-series} yields: \[
\chi(\theta_{c})=\frac{\mu_{B}^{2}}{2\theta_{c}}\]
so that \[
\frac{\chi(\theta=0)}{\chi(\theta_{c})}=\frac{2\sqrt{2}\theta_{c}}{3\pi J\xi_{0}}=\frac{1}{2\sqrt{2}\pi\xi_{0}}<1.\]
The susceptibility at the critical temperature $\theta_{c}$ is continous
as well in agreement with results of Ref. \cite{Hayashi and Ogata }
as they are with the results of the saw-approximation for the spectrum.

\paragraph{High-temperature estimate for the temperature dependence of the susceptibility
in the longitudinal s-RVB state}

At high temperatures (in the vicinity of the upper critical temperature)
the temperature dependent bandwidths of the quasiparticle spectrum
scaling the arguments of the functions in the series eq. (\ref{eq:SusceptibilyExact})
are small. Nevertheless the expansions of the modified Bessel function
and the Struve function at small values of argument cannot be applied
since the summation extends to arbitrary large values of $n$. For
that reason in order to derive temperature dependence of the susceptibility
in the vicinity of the upper critical point we apply the high-temperature
expansion for the energy derivative of the Fermi distribution eq.
(\ref{eq:FermiDistributionDerivative}) as described in Appendix \ref{sub:High-temperature-expansion-for}
and arrive to the formula:\begin{equation}
\chi(\theta)=\frac{\mu_{B}^{2}}{2\theta}\left[1+4\sum_{n=1}^{\infty}\left(\frac{y}{2\theta}\right)^{n}\frac{\left(2^{n+2}-1\right)B_{n+2}}{\left(\frac{n}{2}!\right)^{2}(n+2)}\right],\label{eq:SusceptibilityHighTemperature}\end{equation}
to be used for numerical estimates.

\paragraph{Asymptotic estimate for the temperature dependence of the susceptibility
in the longitudinal s-RVB state}

In order to eluciate the details of the temeperature behavior of the
susceptibility in the longitudinal s-RVB state close to the zero temperature
(provided this state exist in this temperature region) and eventually
valid above the lower critical value we peformed analysis using asympthotic
expansions for the Bessel and Struve functions at large values of
their arguments (low temperatures) as described in Appendix \ref{sub:Asymptotic-estimates-of}
and finally arrived to:\begin{equation}
\chi=\frac{2\mu_{B}^{2}}{\pi y}\left(1+\frac{1}{\sqrt{2\pi}}\sum_{k=1}^{\infty}\frac{\Gamma(\frac{1}{2}+k)}{\Gamma(k+1)}\left(\frac{\pi\theta}{y}\right)^{2k}\left(2^{2k}-2\right)\left|B_{2k}\right|\right)\label{eq:AsymptoticSusceptibility}\end{equation}
The performed analysis is valid provided the ratio $\frac{\xi(\theta)}{\theta}$
has nonvanishing limit or diverges while $\theta\rightarrow0$. In
this case the arguments of the Bessel function in eq. (\ref{eq:SusceptibilyExact})
remain large and the asymptotic expansion used in Appendix \ref{sub:Asymptotic-estimates-of}
on which the final result is based remains valid.

\subsubsection{Two-dimensional s-RVB state}

In the region below the lower critical temperature $\theta_{c}^{\prime}$
both order parameters $\xi$ and $\eta$ of the simplified model are
nonvanishing and the spectrum has a finite gap for all values of the
wave vector in the BZ. Thus the susceptibility can be fairly approximated
by a simple expression:\[
\chi=\frac{2\mu_{B}^{2}}{\theta}\frac{\exp(-\frac{\Delta}{\theta})}{\left[\exp(-\frac{\Delta}{\theta})+1\right]^{2}}\]
where $\Delta$ is some characteristic value of average gap (constant
in the BZ) effectively taking into account both the gap and the dispersion
of the quasiparticles. This expression describes exponentially decaying
susceptibility provided $\frac{\Delta(\theta)}{\theta}\rightarrow\infty$
while $\theta\rightarrow0$. This is in qualitative agreement with
the numerical results of Ref. \cite{Hayashi and Ogata }. Further
improvements can be expected from taking into account temperature
dependence of the average gap below the lower critical temperature
due to expected temperature dependence of the transversal and longitudinal
order parameters.

\section{Comparison with numerical experiment\label{sec:Comparison-with-experiment}}

It has been mentioned above that the precise temperature behavior
of the physical quantities as coming from the analysis of the exact
spectrum or with a saw-like model depend on the details of the temperature
dependence of the order parameters $\xi$ and $\eta$. The numerical
results of Ref. \cite{Hayashi and Ogata } show that the both order
parameters remain finite at $\theta=0$. The high-temperature expansion
as yielding the critical temperatures for the emergence of the both
order parameters yields a linear evanescence of $\xi$ in this limit
in contradiction with results of numerical experiment and with the
exact result eq. (\ref{eq:LimitOrderParameter}). Obviously the high-temparature
expansion fails at low temperatures. On the other hand close inspection
of the numerical results of Ref. \cite{Hayashi and Ogata ,Hayashi-private-communication}
shows that in addition to the vertical tangent characteristic to the
$\xi(\theta)$ dependency at the upper critical temperature one has
to count to visibly horisonthal tangent at zero temperature. This
indicates that the exponent $\alpha$ in eq. (\ref{eq:GeneralizedClassical})
must be larger than unity. With these precautions we shall establish
relation between our analytical results and those of the numerical
modeling in Ref. \cite{Hayashi and Ogata }.

\subsection{Heat capacity}

\begin{figure}
\includegraphics{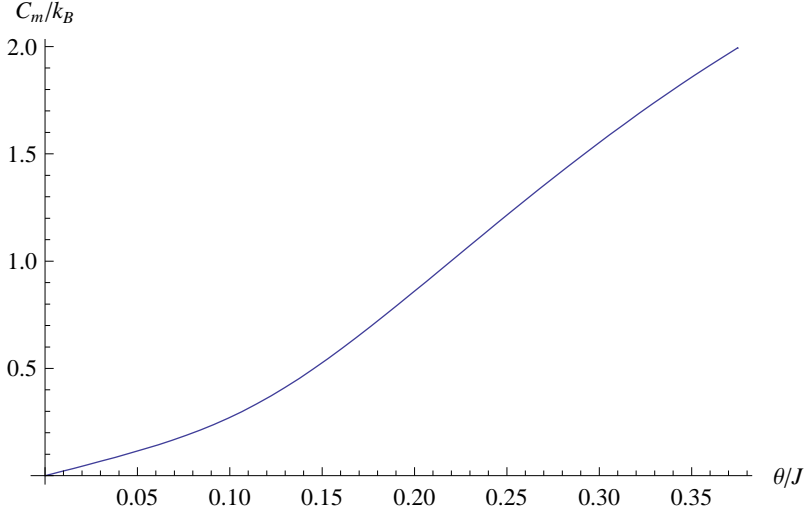}

\caption{$C_{m}$ $vs$ $\theta$ as obtained by numerical integration with
the exact spectrum and with generalized classical approximation for
$\xi(\theta)$ ($\alpha=\pi^{2}/4$).}

\label{fig:Sketch-of-Cm-exact-classical} 
\end{figure}

According to calculations of Ref. \cite{Hayashi and Ogata } the magnetic
contribution to the specific heat capacity experience a jump of $2k_{B}$
in the upper critical point. Combining this with our result eq. (\ref{eq:ExactHeatCapacityJump})
we can derive $\alpha=\pi^{2}/4\approx2.46$. Corresponding graph
represented in Fig. \ref{fig:Sketch-of-Cm-exact-classical}%
shows perfect agreement with the numerical results of Ref. \cite{Hayashi and Ogata }.

\subsection{Magnetic susceptibility}

\begin{figure}
\includegraphics{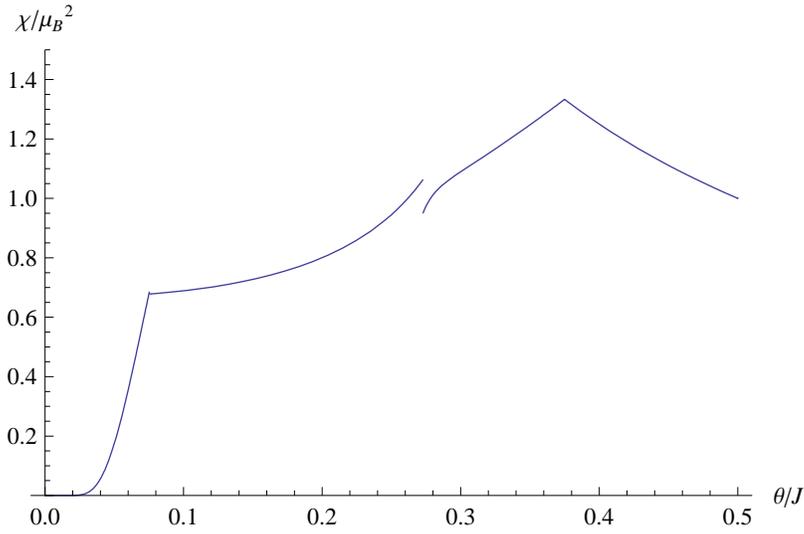}

\caption{$\chi$ \emph{vs} $\theta$ with generalized classical approximation
for $\xi(T)$ ($\alpha=\pi^{2}/4$). and in . The susceptibility in
the low-temperature phase (2D-RVB with both longitudinal and transversal
order parameters nonvanishing) is drawn in the average gap approximation
with the exact magnitude of the latter adjusted to assure continuous
behavior of the susceptibility at the lower critical temperature.
The interval of $\theta/J$ between $\theta_{c}^{\prime}$ and $\frac{3}{11}J$
is drawn according to eq. (\ref{eq:AsymptoticSusceptibility}) with
$k_{max}=1$. The interval of $\theta/J$ between $\frac{3}{11}J$
and $\theta_{c}$ is drawn according to eq. (\ref{eq:SusceptibilityHighTemperature})
up to 14-th power in reciprocal temperature.}

\label{fig:Sketch-of-susceptibility} 
\end{figure}

The results of the model calculation on magnetic susceptibility per
spin are represented in Fig. \ref{fig:Sketch-of-susceptibility} for
the generalized classical approximation for $\xi(\theta)$. 

The temperature dependence of the suceptibility outside the longitudinal
RVB region ($\theta<\theta_{c}^{\prime},\,\theta>\theta_{c}$) is
given respectively by the average gap model and by paramegnetic susceptibility.
Within the longitudinal RVB region ($\theta>\theta_{c}^{\prime},\,\theta<\theta_{c}$)
the asymptotic expansion eq. (\ref{eq:AsymptoticSusceptibility})
is used in the vicinity of the lower critical temperature and the
high-temperature expansion eq. (\ref{eq:SusceptibilityHighTemperature})
in the vicinity of the upper critical point. The picture fairly shows
the key features of the temperature dependence of the magnetic susceptibility
of the longitudinal s-RVB state as they appear from the numerical
calculation of Ref. \cite{Hayashi and Ogata }. The low-temperature
limit is almost achieved pretty above the final exponential drop to
zero characteristic for the low temperature 2D-RVB phase. The asymptotitc
and high-temperature expansions for the susceptibility of the longitudinal
s-RVB state derived for low temperatures remain fairly valid (on respective
sides) almost until quite arbitrary temperature $\frac{3}{11}J$ with
the acceptable amount of discontinuity between two different expansions.

\section{Conclusion}

In the present paper we succeeded in obtaining analytical estimates
for the critical temperatures of the installment of the longitudinal
and 2D s-RVB states in the spin-1/2 Heisenberg model on the anisotropic
triangular lattice in the Ginzburg-Landau approximation for arbitrary
value of the anisotropy (in the assumption of the existence of the
both longitudinal and 2D states). These estimates are in a fair agreement
with the numerical results of Ref. \cite{Hayashi and Ogata } related
to the same system, but apply for the arbitrary anisotropy value $\frac{J^{\prime}}{J}$
in the allowed range. The method does not allow for obtaining correct
estimates of the temperature dependence of the equilibrium values
of the order parameters of the system. That of the longitudinal s-RVB
state was interpolated with use of the generalized classical temperature
dependence corresponding to the critical exponent 1/2 and scaled so
that its value at zero temperature equals to that obtained by minimization
of the zero temperature free energy. 

For the longitudinal s-RVB state of the spin-1/2 Heisenberg model
on the anisotropic triangular lattice, existing between the upper
and lower critical temperatures the analytical estimate of the magnetic
contribution to the specific heat capacity is obtained with use of
the saw-like approximation for the quasiparticle spectrum. The low-temperature
behavior of the magnetic contribution to the specific heat capacity
coincides with that obtained numerically in Ref. \cite{Hayashi and Ogata }:
it has a finite slope at $\theta\rightarrow0$. On the other hand
the magnitude of the abrupt jump of the magnetic specific heat capacity
in the paramagnetic state (above the upper critical temperature) had
been derived for the exact spectrum in agreement with general theory.
In the generalized classical approximation of the temperature dependence
of the order parameter its magnitude ($2k_{B}$) known from numerical
experiment Ref. \cite{Hayashi and Ogata } can be reproduced under
setting the $\alpha$ parameter of the generalized classical formula
equal to $\pi^{2}/4$. 

For the longitudinal s-RVB state of the spin-1/2 Heisenberg model
on the anisotropic triangular lattice analytical expressions are obtained
for the magnetic suceptibility for the saw-like model of the quasiparticle
spectrum as well as for the exact spectrum. (In the latter case --
the asymptotic estimates have been obtained valid respectively in
the vicinity of the lower and higher critical temperatures). Under
assumption of the generalized classical temperature dependence of
the longitudinal order parameter as the most realistic from the point
of view of reproducing the results of numerical experiment on the
magnetic contribution to the specific heat capacity the temperature
dependence of the magnetic susceptibility in the longitudinal s-RVB
state are fairly reproduced: the finite limit at the zero temperature
(under condition of the existence of the longitudinal s-RVB state
at this temperature) is obtained as well as the decrease of the susceptibitity
in the longitudinal phase as compared to the maximal value achieved
at the upper critical temperature, separating the longitudinal s-RVB
phase from the paramegnetic one. Narrow problematic region where both
the low-temperature asymptotic and the high-temperature expansions
are not good enough is found around $\theta\approx3J/11$.

\section*{Acknowledgments}

This work has been performed with the support of Deutsche Forschungsgemeinschaft
and the Excellence Initiative of the German federal and state governments.
In addition, we acknowledge the Russian Foundation for Basic Research
for the financial support dispatched to ALT through the grant No.
10-03-00155. The authors are thankful to Prof. M. Ogata and Mr. Y.
Hayashi for valuable discussion and valuable information which helped
to improve the paper. 

\appendix

\section{Derivation of equations of motions\label{sec:Derivation-of-equations}}

\subsection{Equations of motion for Fermi operators}

The equation of motion for the Fermi operators as coming from the
Heisenberg spin Hamiltonian eq. (\ref{eq:Hamiltonian}) are based
on the commutation relations between the Fermi operators and the spin
operators as given by eq. (\ref{eq:SpinThroughFermi}). Inserting
the explicit form of the tensor components of the spin operator:\begin{eqnarray*}
S_{i}^{+} & = & c_{i\alpha}^{+}c_{i\beta};S_{i}^{-}=c_{i\beta}^{+}c_{i\alpha};\\
S_{i}^{z} & = & (c_{i\alpha}^{+}c_{i\alpha}-c_{i\beta}^{+}c_{i\beta})/2\end{eqnarray*}
 one obtains: \begin{eqnarray*}
[c_{i\alpha},\mathbf{S}_{i}\mathbf{S}_{j}] & = & (c_{i\beta}S_{j}^{-}+c_{i\alpha}S_{j}^{z})/2\\
{}[c_{i\beta},\mathbf{S}_{i}\mathbf{S}_{j}] & = & (S_{j}^{+}c_{i\alpha}-c_{i\beta}S_{j}^{z})/2\end{eqnarray*}

\noindent with this we obtain in terms of the Fermi operators:\begin{eqnarray*}
[c_{i\alpha},\mathbf{S}_{i}\mathbf{S}_{j}] & = & (c_{i\beta}c_{j\beta}^{+}c_{j\alpha}+c_{i\alpha}(c_{j\alpha}^{+}c_{j\alpha}-c_{j\beta}^{+}c_{j\beta})/2)/2\\
{}[c_{i\beta},\mathbf{S}_{i}\mathbf{S}_{j}] & = & (c_{j\alpha}^{+}c_{j\beta}c_{i\alpha}-c_{i\beta}(c_{j\alpha}^{+}c_{j\alpha}-c_{j\beta}^{+}c_{j\beta})/2)/2\end{eqnarray*}
 which must be inserted for each terms in eq. (\ref{eq:Hamiltonian}).

\subsection{Mean field decoupling of equations of motion}

The terms contributing to the right hand side of equations of motion
for Fermi operators as derived in the previous Section are all of
the third overall power with respect to these operators. If one is
interested in having an option for superconducting (anomalous) state
the mean-field decoupling has to read as: \[
c_{1}^{+}c_{2}c_{3}\Rightarrow\left\langle c_{1}^{+}c_{2}\right\rangle c_{3}-\left\langle c_{1}^{+}c_{3}\right\rangle c_{2}+c_{1}^{+}\left\langle c_{2}c_{3}\right\rangle \]
 where the superconducting anomalous averages appear. In the RVB state
we are going to take care of it is reasonable to set \[
\left\langle c_{j\alpha}^{+}c_{j\alpha}\right\rangle =\left\langle c_{j\beta}^{+}c_{j\beta}\right\rangle =\frac{1}{2}.\]
 In order to assure the $S_{z}$conservation we set \[
\left\langle c_{i\alpha}c_{j\alpha}\right\rangle =\left\langle c_{i\beta}c_{j\beta}\right\rangle =0\]
 \[
\left\langle c_{\sigma}^{+}c_{-\sigma}\right\rangle =0\]
 After further assuming the translational invariance and effectively
introducing order parameters $\xi_{\mathbf{\boldsymbol{\tau}}}$:
\begin{eqnarray*}
\left\langle c_{i\alpha}^{+}c_{j\alpha}\right\rangle  & = & \left\langle c_{i\beta}^{+}c_{j\beta}\right\rangle =\xi_{j-i}\\
\xi_{\mathbf{\boldsymbol{\tau}}} & = & \left\langle c_{\mathbf{r}+\boldsymbol{\tau}\sigma}^{+}c_{\mathbf{r}\sigma}\right\rangle \end{eqnarray*}
 \begin{eqnarray*}
\left\langle c_{j\alpha}c_{i\beta}\right\rangle  & = & \left\langle c_{i\alpha}c_{j\beta}\right\rangle =\Delta_{ij}=\Delta_{ji}\\
\left\langle c_{j\beta}c_{i\alpha}\right\rangle  & = & \left\langle c_{i\beta}c_{j\alpha}\right\rangle =-\Delta_{ij}=-\Delta_{ji}\end{eqnarray*}
 \begin{eqnarray*}
\left\langle c_{\mathbf{r}+\boldsymbol{\tau}\alpha}c_{\mathbf{r}\beta}\right\rangle  & = & \left\langle c_{\mathbf{r}\alpha}c_{\mathbf{r}+\boldsymbol{\tau}\beta}\right\rangle =\Delta_{\boldsymbol{\tau}}=\Delta_{\boldsymbol{-\tau}}\\
\left\langle c_{\mathbf{r}+\boldsymbol{\tau}\beta}c_{\mathbf{r}\alpha}\right\rangle  & = & \left\langle c_{\mathbf{r}\beta}c_{\mathbf{r}+\boldsymbol{\tau}\alpha}\right\rangle =-\Delta_{\boldsymbol{\tau}}=-\Delta_{-\boldsymbol{\tau}}\end{eqnarray*}
 Inserting these results in the equation of motion and using the Fourier
transforms for the annihilation operators \[
c_{\mathbf{k}\sigma}=\frac{1}{\sqrt{N}}\sum_{\mathbf{r}}\exp(-i\mathbf{kr})c_{\mathbf{r}\sigma}\]
 ($N$ is the number of sites in the crystal) and hermitean conjugate
one for the creation operators we arrive to:

\noindent the final equations of motion of the form:\begin{eqnarray*}
i\hbar\dot{c}_{\mathbf{k}\alpha}=-\frac{3}{2}\sum_{\mathbf{\tau}}J_{\mathbf{\boldsymbol{\tau}}}\xi_{\mathbf{\boldsymbol{\tau}}}\cos(\mathbf{k\tau})c_{\mathbf{k}\alpha} & +\frac{3}{2}\sum_{\mathbf{\tau}}J_{\mathbf{\boldsymbol{\tau}}}\Delta_{\boldsymbol{\tau}} & \cos(\mathbf{k\tau})c_{-\mathbf{k}\beta}^{+}\\
i\hbar\dot{c}_{\mathbf{k}\beta}=-\frac{3}{2}\sum_{\mathbf{\tau}}J_{\mathbf{\boldsymbol{\tau}}}\xi_{\mathbf{\boldsymbol{\tau}}}\cos(\mathbf{k\tau})c_{\mathbf{k}\beta} & -\frac{3}{2}\sum_{\mathbf{\tau}}J_{\mathbf{\boldsymbol{\tau}}}\Delta_{\boldsymbol{\tau}} & \cos(\mathbf{k\tau})c_{-\mathbf{k}\alpha}^{+}\end{eqnarray*}
 where summation over $\mathbf{\boldsymbol{\tau}}$ is extended to
$\pm\tau_{i};i=1\div3$ .

\subsection{Phase relations for order parameters\label{sub:Phase-relations-for}}

As it can be seen from the definitions eq. (\ref{eq:DispersionFunctions})
of the dispersion functions and the requirement that the final form
of the expression eq. (\ref{eq:eigenvalues}) must be the square root
of a sum of square terms referring to each $\boldsymbol{\tau}$ as
is \emph{mutatu mutandis} eq. (12) of Ref.\cite{Hayashi&Ogata-arxiv}:\begin{eqnarray*}
E_{\mathbf{k}}^{2} & =18 & \left[J^{2}\xi^{2}\cos^{2}(\mathbf{k}_{x})+J^{\prime2}\eta^{2}\left(\cos^{2}(\frac{\mathbf{k}_{x}}{2}+\frac{\mathbf{k}_{y}\sqrt{3}}{2})+\cos^{2}(\frac{\mathbf{k}_{x}}{2}-\frac{\mathbf{k}_{y}\sqrt{3}}{2})\right)\right],\end{eqnarray*}
where we set $\xi=\sqrt{2}D_{1}$ and $\eta=\sqrt{2}D_{23}$. It can
be achieved if the cross terms between the contributions with from
different $\boldsymbol{\tau}$'s cancel each other. This can be done
by specific selection of the relative phases and amplitudes of the
six order parameters $\xi_{\mathbf{\boldsymbol{\tau}}}$ and $\Delta_{\mathbf{\boldsymbol{\tau}}}$.
We assume that \[
\Delta_{\mathbf{\boldsymbol{\tau}}}=\eta_{\mathbf{\boldsymbol{\tau}}}e^{i\varphi_{\tau}}.\]
Then the condition for disapperance of the cross terms in the expression
for $E_{\mathbf{k}}^{2}$ reads:\[
\xi_{\mathbf{\boldsymbol{\tau}}}\xi_{\mathbf{\boldsymbol{\tau}^{\prime}}}=-\eta_{\mathbf{\boldsymbol{\tau}}}\eta_{\mathbf{\boldsymbol{\tau}^{\prime}}}\cos(\varphi_{\tau}-\varphi_{\tau^{\prime}})\]
for all unordered pairs $\boldsymbol{\tau},\boldsymbol{\tau}^{\prime}$
(three conditions). They can be nontrivially satisfied $e.g.$ by
the following choice (not unique due to SU(2) symmetry of the problem):\begin{eqnarray*}
\eta_{1} & = & 0;\xi_{1}\neq0\\
\eta_{2} & \neq & 0;\xi_{2}=0\\
\eta_{3} & \neq & 0;\xi_{3}=0\\
\frac{\pi}{2} & = & \varphi_{2}-\varphi_{3}\\
\xi_{1}^{2} & = & 2\xi^{2}\\
\eta_{2}^{2} & = & \eta_{3}^{2}=2\eta^{2}\end{eqnarray*}
which yields the required form of the excitation spectrum in the RVB
phases.

\section{Series summation\label{sec:Convergence-issues}}

In the present work we use asymptotic series as well as series not
uniformly convergent. The sums of the latter are undestood as limits
of the functions they represent on the border of the convergency range.
Examples follow.

\subsection{Using geometric series for summation of divergent series\label{sub:Using-geometric-series}}

The Fermi distribution can be represented as a shifted geometric series:\begin{eqnarray*}
\left[\exp\left(\frac{E}{\theta}\right)+1\right]^{-1} & = & \frac{\exp(-\frac{E}{\theta})}{1+\exp(-\frac{E}{\theta})}=\exp\left(-\frac{E}{\theta}\right)\sum_{n=0}^{\infty}\left(-1\right)^{n}\exp\left(-\frac{nE}{\theta}\right)\\
 & = & \sum_{n=0}^{\infty}\left(-1\right)^{n}\exp\left(-\frac{(n+1)E}{\theta}\right)\end{eqnarray*}
which is the same as \[
\frac{x}{1+x}=x-x^{2}+x^{3}...=\sum_{n=0}^{\infty}(-1)^{n}x^{n+1}\]
with \[
x=\exp\left(-\frac{E}{\theta}\right)\]
This series converges for all $\left|x\right|<1$ or $\theta>0$,
but at $x=1$ it becomes a divergent series \[
\sum_{n=0}^{\infty}(-1)^{n}=1-1+1-1+1...\]
The function on the left hand, which defines the series, is continous
in $x=1$ and the the value of this function is taken as that of the
sum of the divergent series:

\[
\sum_{n=0}^{\infty}(-1)^{n}=\frac{1}{2}.\]

Analogous consideration applies to the derivative of the Fermi distribution
with respect to the argument of the exponent: \[
-\frac{\exp(\frac{E}{\theta})}{\left[\exp(\frac{E}{\theta})+1\right]^{2}}=\sum_{n=0}^{\infty}\left(-1\right)^{n}(n+1)\exp\left(-\frac{(n+1)E}{\theta}\right).\]
Shifting the summation index by unity makes it a series: \[
\frac{x}{\left(1+x\right)^{2}}=-\sum_{n=0}^{\infty}(-1)^{n+1}(n+1)x^{n+1}=-\sum_{n=1}^{\infty}(-1)^{n}nx^{n},\]
which is the derivative of the geometric series: \[
\frac{1}{1+x}=1-x+x^{2}...=\sum_{n=0}^{\infty}(-1)^{n}x^{n}\]
 \[
\left(\frac{1}{1+x}\right)^{\prime}=-\frac{1}{\left(1+x\right)^{2}}=\sum_{n=0}^{\infty}(-1)^{n}nx^{n-1}=\sum_{n=0}^{\infty}(-1)^{n+1}(n+1)x^{n}\]
multiplied by $x$. The series on the right hand side apparently diverges
at $x=1$ where it becomes:

\[
-\sum_{n=1}^{\infty}(-1)^{n}n=-1+2-3+4...\]
However, its sum, being defined as a limiting value of the function
on left hand side, must be: \[
-\sum_{n=1}^{\infty}(-1)^{n}n=\frac{1}{4},\]
which is used throughout.

\subsection{High-temperature expansion for the susceptibility in the longintudinal
s-RVB state\label{sub:High-temperature-expansion-for}}

In order to obtain the high-temperature expansion for the susceptibility
we notice that the derivative of the Fermi distribution eq. (\ref{eq:FermiDistributionDerivative})
upto a numerical coefficient coincides with the generating function
for the Euler polynomials $E_{n}(x)$ AS23.1.1%
\footnote{Hereinafter ASnn.n.nn stands for the formula nn.n.nn of Ref. \cite{AbramowitzStegun}.%
}:\[
\frac{\partial f}{\partial\varepsilon}=\sum_{n=0}\left(\frac{1}{\theta}\right)^{n+1}\frac{E_{n+1}(0)}{2n!}\varepsilon^{n},\]
which is a good series expansion for small values of $\varepsilon/\theta$
(higher temperatures/small quasiparticle bandwidth). Inserting this
and the exact quasiparticle spectrum $\varepsilon(\mathbf{k}_{x})$
of the longitudinal s-RVB state in the definition of the magnetic
susceptibility eq. (\ref{eq:SusceptibilityDefinition}) we obtain
with account of the integration trick over the Brillouine zone in
the longitudinal s-RVB state Appendix \ref{sec:Integration-in-the}
for the susceptibility:\[
\chi=\frac{2\mu_{B}^{2}}{\pi}\sum_{n=0}^{\infty}\left(\frac{1}{\theta}\right)^{n+1}y^{n}\frac{E_{n+1}(0)}{2n!}\int_{-\frac{\pi}{2}}^{\frac{\pi}{2}}\cos^{n}xdx.\]
Inserting the values of integrals which are $\frac{\sqrt{\pi}\Gamma(\frac{n}{2}+\frac{1}{2})}{\Gamma(\frac{n}{2}+1)}$
and employing the equalities for the factorials and Gamma functions
AS6.1.18 we arrive to:\[
\chi=\mu_{B}^{2}\sum_{n=0}^{\infty}\left(\frac{1}{\theta}\right)^{n+1}\left(\frac{y}{2}\right)^{n}\frac{E_{n+1}(0)}{\left[\left(n/2\right)!\right]^{2}}.\]
Then using AS23.1.20 we get \[
\chi=\frac{\mu_{B}^{2}}{2\theta}\left[1+4\sum_{n=1}^{\infty}\left(\frac{y}{2\theta}\right)^{n}\frac{(2^{n+2}-1)B_{n+2}}{\left[\left(n/2\right)!\right]^{2}(n+2)}\right]\]
(here $B_{n}$ are the Bernoulli numbers) to be used for numerical
estimates.

\subsection{Asymptotic low-temperature expansion of the susceptibility in the
longintudinal s-RVB state\label{sub:Asymptotic-estimates-of}}

We make use of the asymptotic series AS12.2.6for the difference of
the Bessel and Struve functions:\[
I_{0}\left(x\right)-L_{0}\left(x\right)\sim\frac{1}{\pi}\sum_{k=0}^{\infty}\frac{(-1)^{k}\Gamma(\frac{1}{2}+k)}{\Gamma(\frac{1}{2}-k)\left(\frac{x}{2}\right)^{2k+1}}\]
valid for large values of argument:\[
x=\frac{n3\sqrt{2}J\xi}{\theta}=n\frac{y}{\theta}\]
\emph{i.e.} for the lower temperatures. It is remarkable that the
expansion behaves better for larger values of $n$. Inserting this
expansion in the series for the susceptibility we get:\[
\chi=\frac{2\mu_{B}^{2}}{\pi\theta}\sum_{n=1}^{\infty}\sum_{k=0}^{\infty}\left(-1\right)^{n+1}n\frac{(-1)^{k}\Gamma(\frac{1}{2}+k)}{\Gamma(\frac{1}{2}-k)\left(\frac{ny}{2\theta}\right)^{2k+1}},\]
 which results in:\[
\chi=\frac{2\mu_{B}^{2}}{\pi\theta}\sum_{n=1}^{\infty}\sum_{k=0}^{\infty}\left(-1\right)^{n+1}\frac{(-1)^{k}\Gamma(\frac{1}{2}+k)}{\Gamma(\frac{1}{2}-k)n^{2k}}\left(\frac{2\theta}{y}\right)^{2k+1}.\]
 The summation over $n$ then performs with use of AS23.2.19

\[
\sum_{n=1}^{\infty}(-1)^{n+1}n^{-k}=\left(1-2^{1-k}\right)\zeta(k),\]
 where $\zeta(k)$ is the Riemann zeta function, which results in:\[
\chi=\frac{2\mu_{B}^{2}}{\pi\theta}\sum_{k=0}^{\infty}\frac{(-1)^{k}\Gamma(\frac{1}{2}+k)}{\Gamma(\frac{1}{2}-k)}\left(\frac{2\theta}{y}\right)^{2k+1}\left(1-2^{1-2k}\right)\zeta(2k),\]
 where the term with $k=0$ yields: \[
\sum_{n=1}^{\infty}\left(-1\right)^{n+1}n\frac{2\theta}{ny}=\frac{\theta}{y},\]
 according to Section \ref{sub:Using-geometric-series} so that:\[
\chi=\frac{2\mu_{B}^{2}}{\pi y}+\frac{2\mu_{B}^{2}}{\pi\theta}\sum_{k=1}^{\infty}\frac{(-1)^{k}\Gamma(\frac{1}{2}+k)}{\Gamma(\frac{1}{2}-k)}\left(\frac{2\theta}{y}\right)^{2k+1}\left(1-2^{1-2k}\right)\zeta(2k),\]
 where the summation index $k$ is shifted by unity. Next following
AS23.2.16 we get \[
\chi=\frac{2\mu_{B}^{2}}{\pi y}+\frac{\mu_{B}^{2}}{\pi\theta}\sum_{k=1}^{\infty}\frac{(-1)^{k}\Gamma(\frac{1}{2}+k)(2\pi)^{2k}}{(2k)!\Gamma(\frac{1}{2}-k)}\left(\frac{2\theta}{y}\right)^{2k+1}\left(1-2^{1-2k}\right)\left|B_{2k}\right|,\]
 where $B_{2k}$ is the corresponding Bernoulli number.

Further simplification is performed with use of the formulae for the
Gamma function. Using AS6.1.17:

\[
\Gamma(z)\Gamma(1-z)=\frac{\pi}{\sin(\pi z)}\]
 for $z=\frac{1}{2}+k$ we obtain\begin{eqnarray*}
\Gamma(\frac{1}{2}-k) & = & \frac{\pi}{\Gamma(\frac{1}{2}+k)\sin(\pi(\frac{1}{2}+k))};\\
\frac{1}{\Gamma(\frac{1}{2}-k)} & = & \frac{(-1)^{k}\Gamma(\frac{1}{2}+k)}{\pi}\end{eqnarray*}
 so that \[
\chi=\frac{2\mu_{B}^{2}}{\pi y}+\frac{\mu_{B}^{2}}{\pi^{2}\theta}\sum_{k=1}^{\infty}\frac{\Gamma^{2}(\frac{1}{2}+k)(2\pi)^{2k}}{(2k)!}\left(\frac{2\theta}{y}\right)^{2k+1}\left(1-2^{1-2k}\right)\left|B_{2k}\right|\]
Inserting the definition of factorial in terms of Gamma function:
$(2k)!=\Gamma(2k+1)$ and applying the duplication formula AS6.1.18
\[
(2k)!=\frac{1}{\sqrt{2\pi}}2^{2k+1}\Gamma(\frac{1}{2}+k)\Gamma(k+1),\]
we get:\[
\chi=\frac{2\mu_{B}^{2}}{\pi y}\left(1+\frac{1}{\sqrt{2\pi}}\sum_{k=1}^{\infty}\frac{\Gamma(\frac{1}{2}+k)}{\Gamma(k+1)}\left(\frac{\pi\theta}{y}\right)^{2k}\left(2^{2k}-2\right)\left|B_{2k}\right|\right),\]
to be used for numerical estimates.

\section{Integration in the Brillouin zone of the trigonal lattice\label{sec:Integration-in-the}}

The Brillouin zone (BZ) for the triangular lattice is a symmetric
hexagon with a side $4\pi/3$ with the diagonal parallel to the $\mathbf{k}_{x}$
axis in the reciprocal space. Its area equals to $8\pi^{2}/\sqrt{3}$
as employed in the expressions throughout the paper. The integration
over BZ falls into three parts: 
\begin{itemize}
\item one (1) over the bulk of the BZ: $\intop_{-\frac{\pi}{2}}^{\frac{\pi}{2}}$$d\mathbf{k}_{x}$$\intop_{-\frac{2\pi}{\sqrt{3}}}^{\frac{2\pi}{\sqrt{3}}}d\mathbf{k}_{y}$ 
\item one (2) over stripes $2\times\intop_{\frac{\pi}{2}}^{\frac{2\pi}{3}}$$d\mathbf{k}_{x}$$\intop_{-\frac{2\pi}{\sqrt{3}}}^{\frac{2\pi}{\sqrt{3}}}d\mathbf{k}_{y}$ 
\item one (3) over triangles $4\times\intop_{\frac{2\pi}{3}}^{\frac{4\pi}{3}}$$d\mathbf{k}_{x}$$\intop_{0}^{\frac{2\pi}{\sqrt{3}}}d\mathbf{k}_{y}$ 
\end{itemize}
We notice, however, that in the longitudinal s-RVB state the excitation
spectrum is an even function of $\mathbf{k}_{x}$ which consists of
two branches each symmetric with respect to reflection in the straight
lines $\mathbf{k}_{x}=\pm\pi$ which correspond to the ridges of the
respective branches of the excitation spectrum. For that reason while
integrating any function independent on $\mathbf{k}_{y}$ over $\mathbf{k}_{y}$
over the flaps of the BZ (the values of $\frac{2\pi}{3}\leq\left|\mathbf{k}_{x}\right|\leq\frac{4\pi}{3}$
-- triangmes integration) we see that for each given constant value
of the integrand the length of the integration range corresponding
to this value on one side of the ridge as summed with that on the
other side of the ridge yields precisely $\frac{4\pi}{\sqrt{3}}$
which is the width of the BZ in the $\mathbf{k}_{y}$ direction. For
that reason the integration over $\mathbf{k}_{y}$ in the flaps of
the BZ is equivalent to integration over the rectangle $\frac{2\pi}{3}\leq\mathbf{k}_{x}\leq\pi;\left|\mathbf{k}_{y}\right|\leq\frac{2\pi}{\sqrt{3}}$,
which after completing it with the integration over the rectangle
$\frac{\pi}{2}\leq\mathbf{k}_{x}\leq\frac{2\pi}{3};\left|\mathbf{k}_{y}\right|\leq\frac{2\pi}{\sqrt{3}}$
(stripe integration) becomes equal to the integration over the bulk
of the BZ, Thus the integral over the entire BZ in the longitudinal
RBV state satisfies the condition:\[
\frac{\sqrt{3}}{8\pi^{2}}\int_{BZ}f(E_{\mathrm{\mathbf{k}}})d^{2}\mathbf{k}=\frac{1}{\pi}\int_{-\frac{\pi}{2}}^{\frac{\pi}{2}}f(E_{\mathrm{\mathbf{k}}})d\mathbf{k}_{x}\]
as used in our calculations. 
\end{document}